\documentclass[a4paper,12pt]{article}

\usepackage {amssymb,amsmath,bbm,graphicx}

\def\b{\begin{eqnarray}}

\def\e{\end{eqnarray}}

\def\bn{\begin{eqnarray*}}

\def\en{\end{eqnarray*}}

\def\>{\rangle}

\def\<{\langle}

\title{The Projection Operator Method and the Ashtekar-Horowitz-Boulware Model}
\author{J. Scott Little\footnote{little@phys.ufl.edu}
\\Department of Physics
\\
 University of
Florida\\
 P.O. Box 118440\\
Gainesville, FL 32611-8440}
\date{ }
\begin{document}

\maketitle

\begin{abstract}

Motivated by the recent work of Louko and Molgado, we consider the
Ashtekar-Horowitz-Boulware model using the projection operator
formalism.  This paper uses the techniques developed in a recent
paper of Klauder and Little to overcome the potential difficulties
of this particular model.  We also extend the model by including a
larger class of functions than previously considered and evaluate
the classical limit of the model.

\end{abstract}

\section{Introduction}

The Ashtekar-Horowitz model \cite {Ash} was formulated to mimic a particular property of the Hamiltonian constraint of General Relativity.  In this simple model the constraint of the Hamiltonian system was such that the classical constraint subspace did not project down to all of the configuration space. Using the methods described by Dirac \cite {dirac1}, the constraint of this simple quantum mechanical system was imposed.  It was argued that by requiring the additional condition of normalization of the constraint solutions there is quantum mechanical tunneling into classically forbidden regions.  This model was originally formulated with the configuration space of a sphere.

Later Boulware modified the constraint problem by noting that the curvature of the configuration space plays no role in the analysis and altered the configuration space to a torus - a compact yet globally flat configuration space \cite {Bou}.  In the quantization of the modified model, the additional requirement of the self-adjoint property was imposed on the canonical momentum.  Using this additional criteria, it was shown that no tunneling would occur into
the classically forbidden regions for the physical states.

Recently, Louko and Molgudo investigated this model using techniques
of the refined algebraic quantization program (RAQ) to determine its
physical Hilbert space structure \cite {Lou}.  The methods they employed led to the existence of super-selection sectors in the physical Hilbert space.  The manifestation of those sectors were a result of alterations of their tools to account for solutions of the constraint equations that are stationary (critical) points of the arbitrary function $R(y)$ in the constraint.

Using the projection operator formalism \cite {klauder}, we are able to
ascertain the physical Hilbert space of the Ashtekar-Horowitz-Boulware
(AHB) model with techniques which are closer to the essence of the
Dirac procedure \cite {dirac1} than in those in the previous work
\cite {Lou}. The physical Hilbert space of this model is shown not
to decompose into super-selection sectors.   The previous work \cite {kl} serves as a guide for this present endeavor.  See \cite {klauder} and \cite{wayne} for further examples of the projection operator formalism applied to various models.

This paper is organized as follows:  Section 2 introduces the concept of highly irregular constraints and the quantization of these constraints as in \cite{kl}.  Section 3 provides a brief introduction to the classical AHB model.  Section 4 presents the canonical quantization of the model.  Section 5 deals with constructing the physical Hilbert space using the projection operator formalism.  Section 6 deals with the classical limit of the constrained quantum theory and establishes that the classical limit is the classical theory of the original model.  Section 7 contains a brief conclusion and discussion of potential extensions to the model.

\section{Highly Irregular Constraints}

In constrained dynamics one typically places regularity conditions on
the constraint to insure linear independence.  If we consider $N$ classical constraints, $\phi_a$, $a\in\{1,..,N\}$ the regularity condition can be stated in terms of the rank of the Jacobian matrix of the constraints \cite{gov}

\begin{equation}
\mbox{Rank}\frac{\partial^2 \phi_a}{\partial p^n \partial q_m} \Big |_\Gamma = N,
\end{equation}
where $n\in\{1,...,M\}$, $2M$ is the dimensionality of phase space, and $\Gamma$ is the constraint hypersurface $(\phi_a = 0)$.  If this condition fails, then the constraint (or set of constraints) is called irregular \cite{gov}.

Irregular constraints can appear in following form

\begin{equation}
\phi_a^r , \label {irr}
\end{equation}
where $\phi_a$ is a regular constraint and  $r\in\mathbbm{R}^+ \slash \{1\} $ . In
the literature \cite{gov} the measure of irregularity is based on the
order of the zero on the constraint surface.  For example, (\ref{irr})
is an $r^{th}$ order irregular constraint. We should note that while
the constraints $\phi_a$ and $\phi_a^r$ are equivalent (i.e.  the
constraints generate the same constraint hypersurface), the dynamics
and set of observables associated with each given system are {\it not} necessarily equivalent.

The term {\it highly irregular constraint} refers to a constraint
function that involves both regular and irregular constraints or two
or more constraints of varying order \cite{kl}.  For example, let us
consider the following two constraints:
\begin{equation}
\phi_1 = q(1-q)^2,
\end{equation}
\begin{equation}
\phi_2=(q-3)^2(q-4)^3.
\end{equation}
The first constraint is regular at $q=0$ and irregular at $q=1$ of order $2$. The second coinstraint is irregular at both $q=3$, of order 2, and at $q=4$, of order 3.  Both of these constraints are representative of the class of highly irregular constraints. Since the dynamics as well as observability \cite{kl} of a given system are potentially not the same for regular and irregular constraints, careful consideration must be observed when quantizing such systems.  The projection operator formalism \cite{klauder} seems to provide an appropriate framework to deal with systems with irregular constraints \cite {kl}.

The usual form of the projection operator is given by

\begin{equation}
\mathbbm{E}(\Sigma_a \Phi_a^2\leq \delta^2(\hbar)), \label{proj}
\end{equation}
where $\Sigma_a \Phi_a^2$ is the sum of the squares of the
constraint operator and $\delta(\hbar)$ is a small regularization
factor. The projection operator is then used to extract a subspace
of the unconstrained Hilbert space, ${\cal H}$.  If $\Sigma
\Phi_a^2$ has a discrete isolated $0$ then $\delta$ can be chosen to
be an extremely small number.  However, if $\Sigma \Phi_a^2$ has a
$0$ in the continuum, we can not choose an appropriate $\delta$ to
select the proper subspace.  We will discuss this distinct
possiblity shortly. In the limit as $\delta\rightarrow 0$ if
appropriate, this subspace becomes the Physical Hilbert space,

\b
\lim_{\delta\to 0} \mathbbm{E}|\psi\> &\equiv& |\psi\>_{Phys}, \\
\lim_{\delta\to 0} \mathbbm{E}{\cal H} &\equiv& {\cal H}_{Phys}.
\e

 However, if the constraint's spectrum contains a zero in the continuum then the projection operator vanishes as $\delta \rightarrow 0$  \cite{klauder}, which is unacceptable.  To overcome this obstacle, this limit must be evaluated as a rescaled form limit.  To accomplish this, we will need to introduce suitable bras and kets in the unconstrained Hilbert space. For this discussion it will be convenient to choose canonical coherent states ($|p,q\>$) to fulfill this choice.  We regard the following expression as the rescaled form

\begin{equation}
S(\delta)\<p',q'|\mathbbm{E}|p,q\>, \label{form}
\end{equation}
where $S(\delta)$ is the appropriate coefficient needed to extract the
leading contribution of
$\<p',q'|\mathbbm{E}|p,q\>$, for $0<\delta \ll 1$. For example, if
$\<p',q'|\mathbbm{E}|p,q\> \propto \delta$ to leading order, then
$S(\delta)\propto {\delta}^{-1}$, for small $\delta$.   The limit
$\delta\rightarrow 0$ can now be taken in a suitable fashion.  The
expression (\ref{form}) is a function of positive semi-definite type
and this means that it meets the following criteria

\begin{equation}
\lim_{\delta \to 0} \Sigma^N_{j,l=1} \alpha^*_j\alpha_l S(\delta)\<p_j,q_j|\mathbbm{E}|p_l,q_l\> \ge 0,
\end{equation}
for all finite $N$, arbitrary complex numbers $\{\alpha_l\}$ and
coherent state labels $\{p_l,q_l\}$.  A consequence of the previous
statement is that (\ref{form}) can lead to a  reduced reproducing
kernel for the physical Hilbert space

\begin{equation}
{\cal K}(p',q';p,q)\equiv \lim_{\delta \to 0}
S(\delta)\<p',q'|\mathbbm{E}|p,q\>.
\end{equation}

The reproducing kernel completely defines the physical Hilbert space \cite {klauder}.  The reproducing
 kernel makes it possible to express a dense set of vectors in the functional constraint subspace as
\begin{equation}
\psi_P(p,q )= \Sigma^N_{n=1} \alpha_n {\cal{K}}(p,q;p_n,q_n), \hskip1cm N<\infty.
\end{equation}
The inner product for these vectors is given by
\begin{equation}
(\psi,\eta)_P =\sum^{N,M}_{n,m=1}\alpha_n^*\beta_m{\cal
K}(p_n,q_n;p_m,q_m),
\end{equation}
where $\eta$ is also an element of the dense set of vectors.
The completion of these vectors will yield the physical Hilbert space.
Without explicitly calculating the reproducing kernel, we will consider the following highly irregular quantum constraint
\begin{equation}
\Phi= Q^2(1-Q),
\end{equation}
 where $Q$ acts as a multiplication operator.  Clearly this constraint vanishes when $Q=0$ and $Q=1$.   Assuming, $0<\delta\ll 1$, the projection operator for this constraint can be written in the following form
\begin{equation}
\mathbbm{E}(-\delta <\Phi <\delta)=\mathbbm{E}(-\delta <Q^2 <\delta)+\mathbbm{E}(-\delta < (1-Q) <\delta).
\end{equation}
Since the zeros of this operator fall in the continuum, it is clear from the previous discussion we cannot take the limit $\delta \rightarrow 0$ in its present naked form.  The reproducing kernel can be expressed as the following

\begin{equation}
{\cal K}_\diamondsuit = \<p',q'|\mathbbm{E}(-\delta <Q^2 <\delta)|p,q\> + \<p',q'|\mathbbm{E}(-\delta <Q-1 <\delta)|p,q\>.
\end{equation}
By construction these projection operators $\mathbbm{E}(-\delta <Q^2
<\delta)$ and $\mathbbm{E}(-\delta <Q-1 <\delta)$ project onto
orthogonal spaces. To leading order in $\delta(\hbar)$ the reproducing kernel can be approximated by
\begin{equation}
{\cal K}_\diamondsuit \simeq \delta^{1/2}{\cal K}_{Q=0}+ \delta {\cal K}_{Q=1},
\end{equation}
where ${\cal K}_{Q=0}$ and  ${\cal K}_{Q=1}$ are leading order
contributions to the reproducing kernels around the two solutions
to the constraint equation.  See \cite {kl} for further details.  Unlike expression (\ref {form}) there does not exist a single $S(\delta)$ to extract the leading order dependency for the entire Hilbert space.  To address this  difficulty we will consider the following argument \cite {kl}.

Our previous example had constraint solutions around $Q=0$ and $Q=1$, we will now address this in a more general setting.  We begin by determining the reproducing kernel for each solution in the constraint equation.  Recall that the sum of reproducing kernels will produce a direct sum of the corresponding reproducing kernel Hilbert spaces if the spaces are mutually orthogonal.  This will be the case
for highly irregular constraints.  So let ${\cal K}$ represent the
($\delta> 0$) reproducing kernel for the reproducing kernel Hilbert space ${\cal H}$

\begin{equation}
{\cal K} =\Sigma^N_{n=1} {\cal K}_n,
\end{equation}
where ${\cal K}_n$ is the determined reproducing kernel for each unique solution of the constraint.  The Hilbert space generated has the following form,

\begin{equation}
{\cal H} =\bigoplus_{n=1}^N {\cal H}_n,
\end{equation}
where ${\cal H}_n$ corresponds to the ${\cal K}_n$ for each $n$. However, we have not taken the limit as $\delta\rightarrow 0$, and since the leading order $\delta$ dependency is potentially different for each reproducing kernel ${\cal K}_n$, there does not exist a single $S(\delta)$ that can be used to extract the leading order $\delta$ contribution of each reproducing kernel. To accomplish this task we define a (similarity) transformation $S$,

\bn
S: {\cal K}_n &\mapsto&  {\hat {\cal K}}_n, \\
{\hat {\cal K}}_n &=& S_n(\delta) {\cal K}_n,
\en
where $S_n(\delta)>0$ for all $n$ which leads to
\begin{equation}
 {\hat {\cal K}} = \Sigma^N_{n=1}S_n(\delta) {\cal K}_n. \label{rr}
\end{equation}
The rescaled ${\hat {\cal K}}$ serves as the reproducing kernel for
the Hilbert space ${\hat {\cal H}}$.  Although the inner product of ${\cal H}$ and ${\hat {\cal H}}$ are different the set of functions are identical.   The goal of this little exercise is of course to take a suitable limit $\delta \rightarrow 0$ to yield a function that can serve as a reproducing kernel for the physical Hilbert space.  At this point, we can take such a limit.

\b
{\tilde {\cal K}} &\equiv& \lim_{\delta \to 0} {\hat {\cal K}},\\
{\cal H}_{phys} \equiv {\tilde {\cal H}} &=& \bigoplus^N_{n=1}{\tilde {\cal H}}_n,
\e
where ${\tilde {\cal K}}$ is the reduced reproducing kernel for the physical Hilbert space ${\cal H}_{phys}$.  We
will  apply these techniques in Section 5 to the Ashtekar-Horowitz-Boulware model to determine its physical Hilbert space.

\section{Classical Theory}

The classical system of the Ashtekar-Horowitz-Boulware model is given by the following action,

\begin{equation}
I= \int (p_x\dot{x} + p_y\dot{y} -\lambda C) dt, \label{action}
\end{equation}
where $\lambda$ is a Lagrange multiplier corresponding to the constraint $C$.  The constraint has the following form

\begin{equation}
C\equiv p_x^2-R(y), \label{classcon}
\end{equation}
where the function $R(y)\in C^1(\mathbbm{R})$ and is assumed to be
positive somewhere. When the constraint equation is satified the physical states are limited to the regions of the configuration space where $R(y)\geq
0$. The constraint region in the 4 dimensional phase space will involve a proper
subset of configuration space.  The configuration space of the AHB model is
${\cal C}=\mathbbm{T}^2 \simeq S^1 \times S^1 $.   Note
that the Hamiltonian equals zero in this model to emphasize the role of the constraint.

 The dynamics of this system are given by the following 5 equations of motion.

\bn
{\dot x} = -2\lambda p_x,&\hskip.15cm&{\dot y} =0,\\
{\dot p}_x = 0, &\hskip.25cm&{\dot p}_y = \lambda \frac{dR(y)}{dy},\\
& p_x^2-R(y)=0.&
 \en
 From these equations of motion, we can make some statements on some
 observability properties of this theory.  The dynamical variable $x$
 is gauge dependent for all $p_x$ except for $p_x=0$.  The conjugate
 momentum of $y$ also appears to be gauge dependent if the constraints
 are regular around a given set of $y$ that satisfies the constraint
 equation in the phase space.  With this observation we establish that
 the AHB model is an example of a highly irregular constraint, provided there are critical
 points where ${dR}/{dy}|_{y_m} = 0$ where $y_m$ is a solution to the constraint
equation.

\section{Quantum Dynamics}

We now proceed to canonically quantize the system (\ref{action}).  We
will assume our chosen canonical coordinates are Cartesian ones
suitable for quantization \cite {dirac}.  We then promote the
canonical dynamical variables $(x,y,p_x,p_y)$ to a set of irreducible
self-adjoint operators $(X,Y,P_x,P_y)$.  Conjugate pairs corresponding to compact, periodic spatial components will not obey the standard Heisenberg-Weyl relationship \cite {gonz} because the eigenvalues of the conjugate momentum operators are not continuous but discrete.  Continuing with the canonical quantization procedure, we promote the constraint to a suitable function of self-adjoint operators

\begin{equation}
C\mapsto \hat{C} = P_x^2 -R(Y). \label {quantcon}
\end{equation}
Note, there is no ordering ambiguity for this operator.  We assume the
constraint operator is a self-adjoint operator in the unconstrained
Hilbert space.  We can now implement the quantum constraint using the
projection operator  method.

\section{The Physical Hilbert Space via the Reproducing Kernel}

The projection operator for the Ashtekar-Horowitz-Boulware model is chosen to be
\begin{equation}
\mathbbm{E}(\hat{C}^2\leq \delta^2) =\mathbbm{E}(-\delta \leq \hat{C}\le \delta)=\mathbbm{E}(-\delta < \hat{C}< \delta). \label{proj}
\end{equation}
Since the function $R(y)$ is a continuous function, we must introduce an appropriate set of bras and kets to deal with the subtleties described in section 2.

\subsection{The Torus $\mathbbm{T}^2$}

Before constructing the model with the configuration space of a torus we must determine the correct coherent states to use. We wish to use the coherent states not only for computational ease, but also to determine the classical limit, which will be addressed later in this paper.  The torus is the Cartesian product of 2 circles. It follows that the coherent states for the unconstrained Hilbert space can be written as the direct product of 2 coherent states on different circles.

Coherent states on a circle can be generated by coherent states of a
line with the use of the Weil-Brezin-Zak (WBZ) transform \cite {gonz}.
We shall use $\sf{X}$ and $\sf{Y}$ to denote the characteristic
lengths of the $x$ and $y$ coordinates, respectively.  The WBZ
transform, $T$, is a unitary map from $L^2(\mathbbm{R})$ to
$L^2(S^1\times S^{1*})$, where $S^{1*}$ is the dual to $S^1$.  The
transformation is given by the following

\begin{equation}
(T \psi)(x, k) \equiv \Sigma_{n \in \mathbbm{Z}} e^{i n {\sf{X}}k}\psi(x-n\sf{X})
\end{equation}
where $\psi \in L^2(\mathbbm{R})$, $x\in S^1$, and $k\in S^{1*}$ or stated otherwise $k\in [0, \frac{2\pi}{\sf X})$.  We project a corresponding fiber of $L^2(S^1\times S^{1*})$ onto $L^2(S^1)$ by fixing a value of $k$.  Using the standard canonical coherent states in $L^2(\mathbbm{R})$, it has been shown the coherent states on a circle have the following form $(\hbar = 1)$

\b
\nonumber\eta^{(k)}_{x,p}(x')&=&\frac{1}{\pi^{1/4}}\exp(\frac{1}{2}p(x+ip))\exp(-\frac{1}{2}(x+ip-x')^2)\Theta(i\frac{\sf{X}}{2}(x+ip-x'-ik);\rho_1),\\
&\equiv&\<x'|x,p,k\>
\e
where $\rho_1=\exp(-\frac{\sf{X}^2}{2})$ and
\begin{equation}
 \Theta(z) = \Sigma_{n\in \mathbbm{Z}}\rho^{n^2}e^{2inz}, \label{theta}
\end{equation}
 $|\rho|<1$, is the Jacobi theta function.  These states are not normalized \cite{gonz}.  For each value of k these states satisfy the minimal axioms of generalized coherent states; i.e., a continuous labeling of the states where the label set has
 a topology isomorphic to ${\mathbbm R^2}$ and a resolution of unity \cite{klauder2}.

We can express the coherent states on $\mathbbm{T}^2$ as the following,

\begin{equation}
|x,p_x, k_x;y,p_y,k_y\> = |x,p_x,k_x\> \otimes |y,p_y,k_y\>, \label{tor}
\end{equation}
where $x$, $y$ $\in S^1$ and $k_x\in S^{1*}_x$,and $k_y\in
S^{1*}_y$.  We could choose a value for $k_x$ and $k_y$, as well as
${\sf X}$ and ${\sf Y}$.  However, we leave  these choices arbitrary
to allow for the most general solution.

The construction of the reproducing kernel is based on properties of
the constraint operator as well as the coherent states (\ref{tor}).
The constraint operator and the compactness  of $x$ restrict the
spectrum of its conjugate momentum $P_x$ and thereby of $R(y)$.
Allowed values of $y$ are determined by the following equation

\begin{equation}
R(y)= \left(\frac{2\pi n}{\sf{X}}\right)^2, \hskip.5cm n\in \mathbbm{Z}.\label{R}
\end{equation}

We will proceed with the quantization of this model by implementing the method discussed in Section 2 for each $n$ sector of the theory.  Since we are not choosing a particular  $R(y)$, we will only be discussing the physical Hilbert space in general.  We consider the following two types of solutions to the constraint equation.

I.) (Point Solutions)  The solution $y=y_m$ is a point value
solution to the equation (\ref{R}) for a given value of $n$.  The
index $m$ corresponds to multiple values of the $y$ that satisfies
the equation for a given value of $n$. For example, if $R(y)=y^2$,
then for every $n>0$ there exist 2 solutions to the equation
$(\frac{2\pi n}{\sf{X}})^2 =y^2$.

II.) (Interval Solutions)  The solutions $y=y_{m'}$ satisfy the equation (\ref{R}) for all elements
in an interval $I(m')$.  This classification of solutions also includes a countable union of disjoint intervals.
Although physically motivated models exclude such constraint solutions, we include them to illustrate the versatility of our approach
\begin{equation}
\{y_{m'}\} =\{y_{m'}| R(y_{m'}) =(\frac{2\pi n}{\sf{X}})^2 \hskip.5cm \forall y_{m'}\in I_{m'}\}.
\end{equation}

To allow for the greatest generality, it is assumed $R$ will contain both types of solutions.  The calculation of the reproducing kernel can be decomposed into portions corresponding to each value $n\in \mathbbm{Z}$ in the following manner.

Each point solution can be calculated as the following:

\bn
 & & {\cal K}_m(x',p_x', k_x,y',p_y',k_y;x,p_x, k_x;y,p_y,k_y)|_{n=\mbox{constant}}\\
 &\equiv&  \<x',p_x', k_x;y',p_y',k_y|\mathbbm{E}(-\delta \leq R(y)- \left(\frac{2\pi n}{\sf{X}}\right)^2 \le \delta )|x,p_x, k_x;y,p_y,k_y\>\\
 &=&  \int^{y_m +1/S_m(\delta)}_{y_m -1/S_m(\delta)}dy'' \<y',p_y',k_y|y''\>\<y''|y,p_y,k_y\>\<x',p_x',k_x|n, k_x\>\<n, k_x|x,p_x,k_x\>
\en
where $1/S_m(\delta)$ is the leading $\delta$ dependency as
described in Section 2.  For small $\delta$, the integral can  be approximated as follows

\b
\nonumber & &{\cal K}_m(x',p_x', k_x,y',p_y',k_y;x,p_x,
   k_x;y,p_y,k_y) |_{n=\mbox{constant}}\\
\nonumber &=&\frac{2\sin (1/S_m(\delta)(y-y'))}{\sf
{X(y-y')}}e^{i(p_xx-p_x'x') /2+ik_x(x'-x) -(k_x-p_x')^2/2+(k_x+p_x)^2/2}\\
\nonumber &\times&  \exp[-(y_m-y)^2/2 -iy_m(p_y'-p_y)-
(y_m-y')^2/2]\\
\nonumber &\times&\Theta^*(i\frac{{\sf{Y}}}{2}(y'+ip_y'-y_n^m-ik_y;\rho_1)\Theta(i\frac{\sf{Y}}{2}(y+ip_y-y_n^m-ik_y;\rho_1)\\
&\times&\exp[-\frac{4\pi^2}{\sf{X}^2}n^2]\exp[2in[\frac{\pi}{\sf{X}}((x'-x)+i(2k_x-p_x'-p_x)],
\e
where $\rho_1 = \exp[-\frac{\sf{Y}^2}{2}]$. Following the prescription
set forth in Section 2, we perform the required similarity
transformation to extract the leading $\delta$ dependency of
the reproducing kernel.

\begin{eqnarray}
\nonumber {\hat {\cal K}}_{m\, n=\mbox
  {constant}}&=&S_m(\delta)\frac{2\sin
  (1/S_m(\delta)(y-y'))}{\sf {X}(y-y')}e^{i(p_xx-p_x'x')/2+ik_x(x'-x)
  -(k_x-p_x')^2/2+(k_x+p_x)^2/2}\\
\nonumber &\times& \exp[-(y_n^m-y)^2/2
-iy^m_n(p_y'-p_y)- (y_n^m-y')^2/2]\\
\nonumber
&\times&\Theta^*(i\frac{\sf{Y}}{2}(y'+ip_y'-y_m-ik_y;\rho_1)\Theta(i\frac{\sf{Y}}{2}(y+ip_y-y_m-ik_y;\rho_1)\\
&\times&\exp[-\frac{4\pi^2}{\sf{X}^2}n^2]\exp[2in[\frac{\pi}{\sf{X}}((x'-x)+i(2k_x-p_x'-p_x)] .
\end{eqnarray}

The limit $\delta \rightarrow 0$ can now be taken in a suitable manner
to determine the reduced reproducing kernel for this portion of the
physical Hilbert space \cite{klauder} which then reads
\b
\nonumber{\tilde {\cal K}}_{m \, n=\mbox
  {constant}}&=& \frac {2}{{\sf {X}}}e^{i(p_xx-p_x'x')/2+ik_x(x'-x)
  -(k_x-p_x')^2/2+(k_x+p_x)^2/2}\\
\nonumber &\times& \Sigma_m \exp[-(y_n^m-y)^2/2
  -iy^m_n(p_y'-p_y)- (y_n^m-y')^2/2]\\
\nonumber
&\times&\Theta^*(i\frac{\sf{Y}}{2}(y'+ip_y'-y_m-ik_y;\rho_1)\Theta(i\frac{\sf{Y}}{2}(y+ip_y-y_m-ik_y;\rho_1)\\
&\times&\exp[-\frac{4\pi^2}{\sf{X}^2}n^2]\exp[2in[\frac{\pi}{\sf{X}}\{(x'-x)+i(2k_x-p_x'-p_x\}]
\e
for each value of $m$.  Each of these reduced reproducing kernel Hilbert spaces is isomorphic to a one-dimensional Hilbert space (i.e. ${\tilde {\cal H}} \approx \mathbbm{C}$). We continue the procedure for each whole number value of $n$ until the maximum allowed value (of $n$) is reached.

The interval solution portion can be determined in the following computation:
\b
\nonumber& & {\cal K}_{m'}(x',p_x', k_x,y',p_y',k_y;x,p_x,
   k_x;y,p_y,k_y)|_{n=const} \\
\nonumber &\equiv & \<x',p_x',
k_x;y',p_y',k_y|\mathbbm{E}(-\delta \leq R(y)- \left(\frac{2\pi n}{\sf{X}}\right)^2\le \delta )|x,p_x,k_x;y,p_y,k_y\>\\
\nonumber &=& \int_{I_{m'}}dy''
\<y',p_y',k_y|y''\>\<y''|y,p_y,k_y\>\<x',p_x',k_x|n, k_x\>\<n,
k_x|x,p_x,k_x\>\\
\nonumber &=&\frac{2}{\sf {X}} e^{i(p_xx-p_x'x')/2+ik_x(x'-x)
  -(k_x-p_x')^2/2+(k_x+p_x)^2/2}\\
\nonumber &\times& \exp[-\frac{4\pi^2}{{\sf{X}}^2}{n}^2]\exp[2in'[\frac{\pi}{\sf{X}}((x'-x)+i(2k_x-p_x'-p_x)]
  \\
 &\times& \int_{I(m')}dy'' \<y',p_y',k_y|y''\>\<y''|y,p_y,k_y\>.
\e

Once this calculation is preformed for all values of $n$, consistent
with (\ref{R}), then we can
write the reproducing kernel for the physical Hilbert space in the following manner
\begin{equation}
{\tilde {\cal K}} = \Sigma_n^{n_{max}}(\Sigma_m {\tilde{{\cal K}}}_m
  +\Sigma_{m'}{\cal K}_{m'})
\end{equation}
Similarly, the physical Hilbert space can be written as
\begin{equation}
{\cal H}_{phys} = \bigoplus_n^{n_{max}}(\bigoplus_m {\tilde {\cal
    H}}_m \oplus \bigoplus_{m'}  {\cal H}_{m'}).
\end{equation}

The support of the reproducing kernel is only in the classically allowed regions.  This implies there is no tunneling into classically forbidden regions as reported by Boulware \cite{Bou}.

\section{Classical Limit}
We must recall the general rule given by diagonal coherent state matrix elements
\begin{equation}
\<p,q|{\cal{O}}|p,q\>=O(p,q;\hbar),
\end{equation}
where $|p,q\>$ are canonical coherent states.  This provides the
connection between an operator $O(P,Q)$ and an associated function
on the classical phase space manifold.  In the limit, $\hbar
\rightarrow 0$, we find this function reduces to the classical
function that corresponds to quantum operator.  This statement can
easily be seen if ${\cal {O}}$ is a polynomial, however, this
condition is not necessary.  This result can be generalized to any
number of phase space variables as will be demonstrated below.

Before evaluating the classical limit of the model, we must discuss the fundamental difference between quantum mechanics on a compact, periodic configuration space and that of an unbounded space.  The conjugate momentum operator ($P_x$) has a discrete spectrum if the configuration space is compact.  Therefore the standard canonical commutation relation
\begin{equation}
[X, P_x] = i\hbar,
\end{equation}
is inappropriate.  To alleviate this problem we consider the ``angle" operator \cite{gonz}
\begin{equation}
U_x=\exp [(i 2\pi x)/\sf{X}].
\end{equation}
This unitary operator acts to actively translate the operator $P_x$ in the following manner
\begin{equation}
U_xP_xU^\dagger_x = P_x - (2\pi\hbar)/\sf{X}.
\end{equation}
As observed in \cite{kl}, the observable part of an operator can always be expressed as
\begin{equation}
{\cal{O}}^E =\mathbbm{E}{\cal{O}}\mathbbm{E},
\end{equation}
where ${\cal{O}}$ is a self-adjoint operator in the unconstrained Hilbert space.

The observable part of the Hermitian combination of $U_x$ and $U^\dagger_x$ is
\begin{equation}
W_x = U_x^EU_x^{\dagger E} = \mathbbm{E}U_x\mathbbm{E}U_x^\dagger\mathbbm{E}.
\end{equation}
By observation, we note
\begin{equation}
W_x= \mathbbm{E}(-\delta < P_x^2 - R({\bf {Y}})< \delta) {\mathbbm{E}}(-\delta < ( P_x - (2\pi\hbar)/\sf{X})^2 - R({\bf {Y}})< \delta).
\end{equation}
These projection operators are acting on mutually orthogonal
subspaces; therefore, the operator is identically zero.  This result
informs us that this is a gauge dependent question which is
consistent with the classical picture.  Recall from Section 3 the
$x$ dynamical variable is gauge independent only when $p_x=0$.
Quantum mechanically, we have posed the question to find a
``physical" wave function that has support on both a gauge
independent sector and gauge dependent sector.  This is impossible.

If we were to examine the same query for the corresponding Hermitian
combination of the ``angle" operator for the $Y$ coordinate, we
would obtain the unit operator.  The classical limit of this
operator is again in complete agreement to the classical theory.  As
we have previously observed the classical dynamical variable $y$ is
always gauge independent.

Now we consider the following quotient to establish the classical limit of the $Y$ ``angle operator" $U_y$
\begin{equation*}
\frac{\<x,p_x,k_x;y,p_y,k_y|\mathbbm{E}U_y\mathbbm{E}|x,p_x,k_x;y,p_y,k_y\>}{\<x,p_x,k_x;y,p_y,k_y|\mathbbm{E}|x,p_x,k_x;y,p_y,k_y\>} \end{equation*}
\begin{equation}
=\exp[\frac{i 2\pi y}{\sf{Y}} - \frac{\pi^2 \hbar}{{\sf {Y}}^2}]\frac{\Theta (\frac{\sf{Y}}{2\hbar}(p_y-k_y\hbar)-\frac{\pi}{2});\exp(-{\sf{Y}}^2/(4\hbar))}{\Theta(\frac{\sf{Y}}{2\hbar}(p_y-k_y\hbar);\exp(-{\sf{Y}}^2/(4\hbar))}.
\end{equation}
As $\hbar \rightarrow 0$ this expression becomes
\begin{equation}
\exp[\frac{i 2\pi y}{\sf{Y}}],
\end{equation}
where $y$ is subject to the condition $R(y)=p_x^2$.  While this expression is imaginary, we can extract from it the classical reduced phase space coordinate $y$.

Now we direct our attention to the expectation value of the physical conjugate momentum, $P_x$

\begin{equation*}
\frac{\<x,p_x,k_x;y,p_y,k_y|\mathbbm{E}P_x\mathbbm{E}|x,p_x,k_x;y,p_y,k_y\>}{\<x,p_x,k_x;y,p_y,k_y|\mathbbm{E}|x,p_x,k_x;y,p_y,k_y\>}
\end{equation*}

\begin{equation}
=\frac{-i\hbar}{{\tilde{{\cal{K}}}}}\int dy' \int dx' \<x,p_x,k_x;y,p_y,k_y|x',y'\>\frac{\partial}{\partial x'}\<x',y'|x,p_x,k_x;y,p_y,k_y\>.
\end{equation}

We implement the constraints by integrating over the appropriate intervals as described in Section 5.  We can continue this calculation in a similar
 manner to that which is performed in \cite{kl}.

\begin{equation}
\frac{\<x,p_x,k_x;y,p_y,k_y|\mathbbm{E}P_x\mathbbm{E}|x,p_x,k_x;y,p_y,k_y\>}{\<x,p_x,k_x;y,p_y,k_y|\mathbbm{E}|x,p_x,k_x;y,p_y,k_y\>} =p_x+ \frac{{\sf{X}}\Theta '(\frac{\sf{X}}{2\hbar}(p_x-k_x\hbar);\exp(-{\sf{X}}^2/(4\hbar))}{4\Theta(\frac{\sf{X}}{2\hbar}(p_x-k_x\hbar);\exp(-{\sf{X}}^2/(4\hbar))},
\end{equation}
where prime $'$ denotes the derivative with respect to $x'$.
As $\hbar$ approaches $0$, the second term vanishes which can be seen in the definition of the Jacobi theta function (\ref{theta}) \cite{gonz}, thus recovering this aspect of the classical theory from its quantum analog.  Using the same technique, we can also calculate the classical limit of the expectation value of the $P_y$ operator.  The projection operator formalism is well suited to not only properly impose quantum constraints but also allow one to return to the proper classical theory in the limit $\hbar \rightarrow 0$.

\section{Conclusions and Summary}

Motivated by the recent work of Louko and Molgado \cite{Lou}, we
have investigated the Ashtekar-Horowitz-Boulware model using the
projection operator formalism.  Unlike the physical Hilbert space
obtained by RAQ methods, the physical Hilbert space obtained by
implementing the projection operator formalism contains  no
superselection sectors.  It is believed that the superselection
sectors found in \cite{Lou} are consequences of the machinery of the
refined algebraic quantization approach.   We were also able to
extend the model by including a larger class of functions $R(y)$
than previously considered for this model.   The methods can be
generalized to include other configurations spaces such as a
cylinder and a plane.  We were also able to show that in the limit
as $\hbar \rightarrow 0$, the classical theory is recovered.  In a
future work we will consider these extensions.
\section{Acknowledgements}
John Klauder is thanked for all of his help and support during this project.


\begin{thebibliography}{99}

\bibitem{Ash} A. Ashtekar and G. T. Horowitz, ``On the canonical approach to quantum gravity", Phys. Rev. D {\bf 26}, 3342 (1982).
\bibitem{dirac1} P.A.M. Dirac, {\it Lectures on Quantum Mechanics,} (Belfer Graduate School of Science, Yeshiva University, New York, 1964).
\bibitem{Bou} D. G. Boulware, ``Comment on `On the canonical approach to quantum gravity' ", Phys. Rev. D {\bf 28}, 414 (1982).
\bibitem{Lou} J. Louko and A. Molgado, ``Superselection sectors in the Ashtekar-Horowitz-Boulware model", Class. Quantum Grav. {\bf 22} 4007-4019.
\bibitem{klauder} J.R. Klauder, ``Quantization of Constrained Systems'', Lect. Notes Phys. {\bf 572}, 143-182 (2001); ``Coherent State Quantization of Constraint Systems'', Ann. Phys. {\bf 254}  419-453 (1997).
\bibitem{kl} J.R. Klauder and J. S. Little, ``Highly Irregular Quantum Constraints'', Class. Quant. Grav. {\bf 23}, 3641 (2006).
\bibitem{wayne} W.R. Bomstad and J.R. Klauder, ``Linearized Quantum Gravity Using the Projection Operator Formalism'', gr-qc/0601087.
\bibitem{gov} J. Govaerts, {\it Hamiltonian Quantisation and Constrained Dynamics} (Leuven University Press, Belgium, 1991); D.M. Gitman and I.V. Tyutin, {\it Quantization of Fields with Constraints} (Springer-Verlag, Berlin, Heidelberg 1990); M. Henneaux and C. Teitelboim, {\it Quantization of Gauge Systems}, (Princeton University Press, NJ, 1992). See also O. Mi\u skovi\'c and J. Zanelli, ``Dynamical Structure of Irregular Constrained Systems'', J. Math. Phys. {\bf 44}, 3876-3887 (2003).

\bibitem{dirac} P.A.M Dirac, {\it The Principles of Quantum Mechanics}, (Oxford Science Publications, 1999), 4th Edition, p. 114.
\bibitem{gonz} J. A Gonz\'alez and M. A. del Olmo, ``Coherent States on the circle", J. PHys. A: Math. Gen. {\bf 31} 8841-8857 (1998).
\bibitem{klauder2} J. R. Klauder, B. Skagerstam, {\it Coherent States}, (World Scientific Publishing, 1985).
\end{thebibliography}
\end{document}